# Imaging the emergence of bacterial turbulence: phase diagram and transition kinetics

Yi Peng[1,2], Zhengyang Liu[1] and Xiang Cheng[1,*]

[1] Department of Chemical Engineering and Materials Science, University of Minnesota, Minneapolis, MN 55455, USA

[2] Beijing National Laboratory for Condensed Matter Physics, Institute of Physics, Chinese Academy of Sciences, Beijing 100190, China

[*] Corresponding author: Email: xcheng@umn.edu

We experimentally study the emergence of collective bacterial swimming, a phenomenon often referred to as bacterial turbulence. A phase diagram of the flow of 3D *E. coli* suspensions spanned by bacterial concentration, the swimming speed of bacteria and the number fraction of active swimmers is systematically mapped, which shows quantitative agreement with kinetic theories and demonstrates the dominant role of hydrodynamic interactions in bacterial collective swimming. More importantly, we trigger bacterial turbulence by suddenly increasing the swimming speed of light-powered bacteria and image the transition to the turbulence in real time. Our experiments identify two unusual kinetic pathways, i.e., the one-step transition with long incubation periods near the phase boundary and the two-step transition driven by long-wavelength instabilities deep inside the turbulent phase. Our study provides not only a quantitative verification of existing theories, but also new insights into interparticle interactions and transition kinetics of bacterial turbulence.

# Introduction

Collective motions of biological systems such as bird flocks, fish schools and bacterial swarms are the most vivid examples of the emergent behaviors of active matter (*1*). While moving independently at low density, self-propelled units in active matter can move collectively at high density, giving rise to coherent flows at length scales much larger than the size of individual units. In bacterial suspensions, these coherent flows exhibit a chaotic pattern of intermittent vortices and jets, reminiscent of turbulent flows at high Reynolds numbers. Hence, the flows induced by bacterial collective swimming are also known as active or bacterial turbulence (*2-5*).

Extensive theoretical and numerical studies have been conducted in understanding the physical principles underlying the nonequilibrium transition between the disordered and the turbulent states in bacterial suspensions (*1, 6-8*). Particularly, kinetic theories have been developed by extending the classic models of suspensions of passive rod-shaped particles (*7-9*). The theories consider the probability distribution of the position and orientation of bacteria based on a Smoluchowski equation, where a bacterium is modeled as a rigid rod exerting a pusher-type force dipole on the suspending fluid. In addition to self-propulsion, the motion of bacteria is further coupled with a mean-field background Stokes flow, driven by an average active dipolar stress as well as more conventional particle-induced viscous stresses. Using the kinetic theories, Saintillan and Shelley first showed that a long wavelength hydrodynamic instability drives the transition to active turbulence in suspensions of pusher swimmers (*8, 10-12*). The transition kinetics was further studied beyond the linear regime by numerically solving kinetic equations (*11*) and simulations of self-propelled rods (*13*). By incorporating the effect of tumbling and rotational diffusion in the kinetic theories, Subramanian and Koch independently showed that the isotropic disordered state of bacterial suspensions becomes unstable above a threshold concentration in the bulk limit (*7, 14*). The threshold concentration increases with the random reorientation and decreases with the swimming speed of bacteria. The kinetic theories generally assume long-range hydrodynamic dipolar interactions between bacteria and are therefore valid only for dilute suspensions. Nevertheless, the onset of bacterial collective swimming is typically observed at intermediate to high concentrations where steric interactions are supposed to be important (*15-17*), which raises the question on the relative importance of hydrodynamic versus steric interactions in the formation of bacterial turbulence. Indeed, simulations and models with steric interactions have also been developed, which successfully predicted the rise of collective motions both with hydrodynamic interactions (*15, 16*) and without (*18, 19*). Although steric interactions have been shown to play a leading role in bacterial collective swimming in 2D or quasi-2D systems such as Hele-Shaw cells with two confining walls (*20, 21*) and bacterial monolayers on agar substrates (*22, 23*), it is still not clear what is the dominant interaction leading to bacterial turbulence in 3D suspensions.

In comparison with the theoretical development, definitive experiments on bacterial suspensions that can quantitatively verify theoretical predictions are still few and far between (*7, 8*). Pioneering experiments on suspensions of *Bacillus subtilis* first showed the emergence of bacterial turbulence at high concentrations in absence of bioconvection and chemotaxis (*17, 24-27*). Various physical properties of bacterial turbulence such as density fluctuations (*27, 28*), coherent flow structures (*28-32*) and mass and momentum transports (*33-36*) have been subsequently studied. While most of these studies focused on the rise of bacterial turbulence with increasing bacterial concentrations, few experiments have considered other factors important for bacterial turbulence such as the swimming speed of bacteria (*29, 30*) and the presence of defective immobile bacteria. A systematic mapping of the phase diagram of 3D bacterial flows over a large control parameter space is still lacking. Such an experimental phase diagram is essential to verify the quantitative prediction of the kinetic theories. Particularly, the study of the effect of doping immobile bacteria on the phase dynamics will provide evidence to resolve the controversy over the dominant interaction responsible for collective swimming in 3D suspensions of microorganisms (*15, 16*). Increasing the ratio of immobile to mobile bacteria in a suspension of a fixed total bacterial concentration would proportionally reduce the active stress and thus suppress the hydrodynamics-driven collective motion. On the contrary, adding immobile bacteria to a suspension of a fixed concentration of mobile bacteria would promote steric interactions due to the rigid rod-shaped body of immobile bacteria and therefore enhance the collective motion if steric interactions were the leading cause of bacterial turbulence.

Moreover, the kinetic pathway towards the collective motion of swimming microorganisms remains largely unexplored. The kinetics of a phase transition affects both the transition rate and the structure of intermediate states and is of importance in understanding the phase dynamics of equilibrium systems (*37*). It plays a key role in distinguishing first-order and second-order phase transitions. Kinetics is equally important in a nonequilibrium phase transition, which can reveal not only the nature of the transition but also the missing link between the rise of the macroscopic order and the microscopic interparticle interaction. Hence, resolving the route to bacterial turbulence—a premier example of collective motions—addresses the central question in active matter: how do random self-propelled units self-organize into large-scale dynamic structures?

Here, we aim to address the above questions. Our experiments provide not only the most comprehensive phase diagram of the flow of 3D bacterial suspensions heretofore, but also the detailed characterization of the kinetic route of the bacterial turbulent transition. Our experimental phase diagram quantitatively agrees with the kinetic theories, as well as a simple hydrodynamic model balancing dipolar interactions and rotational diffusion of bacteria. By further examining the effect of doping immobile bacteria on the phase dynamics, we show that hydrodynamic interactions dominate the formation of collective swimming in 3D bacterial suspensions. Moreover, our kinetic measurements reveal an unexpected kinetic pathway near the phase boundary in analogy of the nucleation and growth process in equilibrium phase transitions and confirm the existence of a long-wavelength instability deep inside the turbulent phase. Taken together, our experiments validate the basic assumption of the kinetic theories and corroborate the principal predictions of the theories on the transition point and the mode of instability. Furthermore, our study uncovers new kinetic features of the bacterial turbulent transition and sheds new light on the nature of the nonequilibrium phase transition in active microbiological systems. These findings open new questions for future experimental and theoretical studies.

**Results**

We use *Escherichia coli* (*E. coli*) as our model bacteria. In order to examine a large control parameter space, three different strains of *E. coli* are used in our experiments, which all share similar body plans (Materials and methods). In addition to a wild-type strain (BW25113), a strain of tumblers with *cheZ* knocked out (RP1616) and flagella removed is used as immobile passive bacteria. Moreover, a strain of light-powered *E. coli* (BW25113 with proteorhodopsin, a light-driven transmembrane proton pump (*38, 39*)) is also used, whose motility can be varied by green light of different intensities (Fig. S1 in Supplementary Materials (SM)). By mixing different strains of bacteria and controlling light intensity, we explore the phase diagram of 3D bacterial flows spanned by bacterial volume fraction $\phi$, the average swimming speed of bacteria $v$ and the number fraction of active swimmers $f$ (Materials and methods).

**Transition to bacterial turbulence.** The collective motion of bacteria leads to bacterial turbulence with characteristic intermittent jets and whirls, whose lengths and speeds are much larger than those of individual bacteria (Figs. 1A, B and Movie S1). To quantify bacterial turbulence, we calculate the orientational correlation of local velocities, $c_i = \min_{j=1..4}(\hat{v}_i \cdot \hat{v}_j)$, where $\hat{v}_i$ is the unit vector along the direction of the local velocity at position *i* and $\hat{v}_j$ is the direction of the velocity in one of its four neighboring boxes adopted in Particle Imaging Velocimetry (PIV). Here, our PIV measures 2D in-plane flows $\vec{v} = (v_x, v_y)$ in 3D bacterial suspensions (Materials and methods). We identify a local region with high velocity orientational correlation when $c \geq 0.9$. The area fraction of these highly correlated regions, *M*, is used in our study as the order parameter quantifying the rise of bacterial turbulence (Fig. 1C). This choice is similar to that used in previous studies (*17*). Quantitatively similar results can also be obtained if the average orientational correlation $\langle c \rangle$, the velocity correlation length or the effective diffusivity of passive tracers embedded in bacterial suspensions is used as the order parameter (Fig. S2). In addition to the velocity orientational order, we also measure the strength of bacterial flows via kinetic energy per unit mass, $E = \langle v_x^2 + v_y^2 \rangle/2$. Although the definition of kinetic energy density is the same as that used for classical turbulent flows, the inertia of bacterial swimming in our experiments is negligible with the Reynolds number on the order of $10^{-4}$.

A transition to bacterial turbulence is observed as we increase bacterial volume fraction, $\phi$ (Fig. 2A). A sharp increase of $M$ occurs around $\phi = 0.018$. The transition point, $\phi_c$, is obtained from the inflection point of the error function fitting of $M(\phi)$. The transition to bacterial turbulence at $\phi_c$ is also coincident with a more gradual increase of $E$ (Fig. 2A). We find similar sharp transitions to bacterial turbulence when increasing $v$ and $f$ at fixed $\phi$ (Figs. 2B, C). Particularly, bacterial turbulence is observed at $f = 20\%$ for the mixture of wild-type *E. coli* ($v = 28.7$ μm/s) and immobile bacteria at $\phi = 0.18$. Such a low $f$ demonstrates the robustness of the collective flow of bacterial suspensions, greatly surpassing the state-of-the-art design of active robotic systems (*40*). This high tolerance to malfunctioning units likely arises from long-range hydrodynamic interactions, lacking in dry active matter (*1*).

**3D phase diagram.** Measurements over a thousand bacterial samples under different combinations of control parameters yield a full 3D phase diagram of bacterial flows, where bacterial turbulence emerges at large $\phi$, $v$ and $f$ (Fig. 2D). Although the transition to bacterial turbulence has been investigated with increasing $\phi$ and $v$ in separate experiments (*17, 26-30*), to the best of our knowledge, systematic measurements over such a large parameter space in the same experimental system have not been achieved previously. This comprehensive 3D phase diagram not only allows us to quantitatively verify the theoretical prediction on the transition point, but also sets up a framework for exploring the kinetics of the transition in the next section.

We estimate the phase boundary between disordered and turbulent phases using a simple hydrodynamic model proposed in Ref. (*41*), where the pairwise hydrodynamic interaction that promotes local bacterial alignment competes with the rotational relaxation that disorientates bacteria. Although these two competing factors also form the basic ingredients of the kinetic theories (*7, 8*), this simple model considers only the interaction between a pair of bacteria, instead of mean-field bacterial probability distributions. Specifically, we calculate the rotation of a bacterium in a flow field induced by the swimming of its neighbor at distance $r$. A critical distance $r_c$ is determined when the flow-driven rotation of the bacterium balances the reorientation of the bacterium due to random tumbling and rotational diffusion. The transition volume fraction is then given as $\phi_c = V_b/r_c^3$, where $V_b = 2\pi ab^2 = 2.20$ μm$^3$ is the volume of a single bacterium. Here, we approximate the bacterial body as a cylinder with an average length $2a = 2.8$ μm and radius $b = 0.5$ μm. The volume of bacterial flagella is too small to be considered when calculating $V_b$. To incorporate the presence of immobile bacteria, we extend the original model and include the higher-order hydrodynamic interaction due to reflection flows in the calculation. The detail of the model derivation can be found in Supplementary Materials. The model predicts the transition concentration $\phi_c$ in the bulk limit

$$\phi_c - \frac{r_0^3}{V_b}\phi_c^2 = \frac{v_0}{fv}, \tag{1}$$

where $r_0 = 1.6$ μm provides a characteristic interaction range of the reflection flows and $v_0$ is a velocity scale determined by the strength of the force dipole and the rotational relaxation of bacteria

$$v_0 = \frac{8\pi\eta V_b}{\xi l(\Gamma+1)}\sqrt{\frac{20D_r}{3\tau}}.$$

Here, $\eta = 1$ mPa·s is the viscosity of suspension buffer, $\xi = 1.91 \times 10^{-8}$ kg/s is the drag coefficient of bacterial body, $l = 1.9$ μm is the distance of the force dipole and $D_r = 0.057$ s$^{-1}$ is the effective diffusion coefficient of *E. coli*. $\xi$, $l$ and $D_r$ are all obtained from direct experimental measurements (*41*). $\Gamma = (a^2-b^2)/(a^2+b^2) = 0.77$ is the shape factor of a bacterial body. Lastly, $\tau$ quantifies the interaction time of the two bacteria bounded between $a/v \sim 0.1$ s and the average tumbling time $\tau_t \sim 1$ s (*43*). Direct fitting of the phase boundary yields $\tau = 0.85 \pm 0.04$ s within the bound, which leads to $v_0 = 0.57$ μm/s. Eq. (1) predicts a collapse of the transition boundaries of different parameters and shows an excellent agreement with experiments (Fig. 2E). The rescaled phase boundary shows that the fraction of active swimmers couples with the velocity of bacteria in determining the transition to bacterial turbulence. The equivalence of $f$ and $v$ can be further seen from the dynamic structure of the turbulent flows beyond the transition point, where the velocity correlation length of bacterial flows at different $v$ and $f$ also collapses into a master curve when plotting against $fv$ (Fig. S2C). Thus, by varying $f$, we can effectively extend bacterial velocity over a broader range in experiments.

The simple hydrodynamic model provides a physically intuitive picture for the emergence of bacterial collective motions. However, to incorporate many-body interactions beyond two bacteria and avoid the divergence associated with the slow algebraic decay of the dipolar interaction, one needs to consider effective media or averaged equations adapted in the kinetic theories (*14*). Thus, we compare our experiments with the prediction of the kinetic theories. First, at low $\phi_c$, the higher order term on the left side of Eq. (1) is negligible, which leads to $\phi_c \sim 1/v$, consistent with the kinetic theories (*7, 9, 42*). Quantitatively, Koch and Subramanian predicted the transition volume fraction in the bulk limit (*7*)

$$\phi_c = \frac{\eta V_b}{\xi l}\left[\frac{5}{\tau_t} + 30D_r\right]\frac{1}{fv}, \tag{2}$$

where $\xi = 6\pi\eta b\left[1 - \frac{1}{5}\left(1 - \frac{a}{b}\right)\right] \approx 1.28 \times 10^{-8}$ kg/s is the calculated drag coefficient of bacterial body orientated along its major axis, which is slightly smaller than the measured value shown above. The linear stability analysis of the original kinetic theories can be easily extended for the mixture of mobile and immobile bacteria (*14, 44*), which results in *f* as the prefactor of *v* in Eq. (2). Without fitting parameters, Eq. (2) quantitatively agrees with our experiments (Fig. 2E), although small deviations between the theory and experiments can be observed at high concentrations. Thus, our experimental phase diagram provides a quantitative verification of the prediction of the kinetic theories on the transition point. A comparison between the simple hydrodynamic model (Eq. (1)) and the kinetic theories (Eq. (2)) is further drawn in SM. Our experiments also confirm numerical results from large-scale particle-based simulations with hydrodynamic interactions and random reorientations (*45, 46*), where collective motions have been observed above the predicted threshold concentration.

At $f = 1$, $\phi_c$ ranges between 0.02 and 0.08 for different bacterial velocities (Fig. 2E inset). The average distance between bacteria at $\phi_c$ is about 6 to 9.5 μm, on the order of the size of a single bacterium. Short-range hydrodynamic and steric interactions could be important at such small scales (*15, 16*). Hence, it is nontrivial that the far-field dipolar hydrodynamic interaction used in the simple hydrodynamic model (Eq. (1)) and in the kinetic theories (Eq. (2)) is sufficient to describe the experimental phase boundary. The observation supports the basic assumption of the kinetic theories (*7, 8*). Consistent with this finding, increasing the fraction of immobile bacteria at either a fixed total bacterial volume fraction $\phi$ or a fixed volume fraction of active swimmers $f\phi$ suppresses bacterial turbulence in our experiments (Fig. S3), rather than promoting the collective motion as observed in dry active matter with purely short-range steric interactions (*47*). Thus, our study provides experimental evidence on the dominant role of long-range hydrodynamic interactions in 3D suspensions, resolving the controversy over the interparticle interaction responsible for 3D bacterial collective swimming. Physically, if bacterial orientation is random, the effective volume occupied by a bacterium should be considered as a sphere of radius $a$ instead of the excluded volume of the bacterial body $V_b$, which would lead to much larger effective $\phi_c$ up to 0.41 in the semi-dilute regime (*17*). The dominant role of the long-range dipolar interaction observed in our experiments suggests that bacterial orientation is strongly correlated near the phase boundary due to the hydrodynamic interaction. As a result, bacteria show preferred alignment with their neighbors, giving rise to relatively large inter-bacteria distances suitable for the long-range hydrodynamic interaction.

Lastly, it is worth noting that the size of samples in our experiments is fixed with the minimal dimension between 150 μm and 170 μm, much larger than the size of single bacteria. The volume of our samples is on the order of several microliters, which contains more than $10^7$ bacteria at $\phi = 0.018$. Simulations and kinetic theories have shown that $\phi_c$ is independent of the system size at such large scales (*9*). More recent experiments have also shown that the onset of collective motions and bacterial superfluids is insensitive to the system size above 170 μm (*48*). Thus, the phase diagram shown in Fig. 2D should approximate the phase behavior of bacterial suspensions in the bulk limit.

**Transition kinetics.** Next, using the phase diagram as a roadmap, we explore the kinetics of bacterial turbulent transition, which reveals the mode of instability and the transient structure of bacterial suspensions through the transition. To study the kinetics, we trigger the onset of bacterial turbulence by suddenly ramping up light intensity at $t = 0$, which increases the swimming speed of light-powered bacteria from that below the phase boundary to high $v$ above the boundary at given $\phi$ and $f$ (Movies S2 and S3). Although individual bacteria in dilute suspensions recover their swimming speeds within a couple of seconds (Fig. S1), the emergence of collective flows can take much longer times depending on the specific control parameters. In the region above but close to the phase boundary, the transition exhibits a surprisingly long incubation period with low velocity orientational order $M$ and kinetic energy $E$ (~ 100 s in Fig. 3A and Movie S2). During incubation, localized regions with relatively large velocity orientational correlation (high $c$) and large kinetic energy (high $E$) nucleate within the disordered phase (Fig. 3B). The incubation period witnesses the slow growth and the fluctuation of these high-$c$ and high-$E$ regions. After incubation, the high-$c$ and high-$E$ regions grow quickly and eventually percolate the entire system. $M$ and $E$ are directly coupled and increase simultaneously during the transition.

The incubation period becomes increasingly short as we move away from the phase boundary. Far above the boundary, the increase of the velocity orientational correlation is almost instantaneous across the entire system after the light ramp (Figs. 3C, D and Movie S3), whereas the kinetic energy remains low at beginning and grows to a steady-state plateau only at a later time. The growths of $M$ and $E$ are decoupled in this case. The system exhibits a transient state with high velocity orientational order but low kinetic energy (the middle two pictures at $t = 10$ and 30 s in Fig. 3D).

A smooth crossover from the single-step transition near the phase boundary to the two-step transition with the transient state deep inside the turbulent phase can be seen from the $E$-$M$ plot (Fig. 4A). $E$ increases linearly with $M$ near the phase boundary at low $\phi$ indicating the concurrent increase of the two quantities, whereas an L-shaped $E$-$M$ relation is observed for the two-step transition at high $\phi$ deep inside the turbulent phase. The fast increase of $M$ at early times results in the flat region of the curves at low $E$.

We further quantify the time and length scales associated with transition kinetics. The transition rate is measured by the incubation time, $\tau_{in}$, defined in analogy of nucleation and growth processes in equilibrium phase transitions (Fig. S4). $\tau_{in}$ decreases with $\phi$ and reaches a low plateau of ~ 10 s above $\phi = 0.105$ (Fig. 4B), comparable to the time it takes for a single bacterium recovering its swimming speed in the dilute limit. $1/\tau_{in}$ provides an estimate the growth rate of the hydrodynamic instability. The presence of the transient state with high velocity orientational correlation and low kinetic energy is characterized by the time difference, $\Delta\tau = \tau_E - \tau_M$, where $\tau_E$ and $\tau_M$ are the times when $E$ and $M$ reach their steady states, respectively. $\Delta\tau$ increases with $\phi$ and can be as large as 65 s above $\phi = 0.16$ deep inside the turbulent phase.

Note that although we discuss transition kinetics with increasing $\phi$, qualitatively similar trends are also observed when $v$ and $f$ are varied (Figs. 4D-F).

The length scales associated with the transition kinetics are revealed by the energy spectrum of bacterial flows, $E(k)$, where $k$ is the wavenumber (*3, 28*) (Materials and methods). $E(k)$ is related to the energy density $E$ through $E = \int_0^\infty E(k)dk$. Near the phase boundary, the increase of $E(k)$ initiates at large $k$ (or short wavelengths) and then propagates to small $k$ over time (Fig. 5A), consistent with the scenario of nucleation and growth. In sharp contrast, deep inside the turbulent phase, the spectrum increases significantly faster at small $k$ at early times (Fig. 5B), indicating a long-wavelength instability. Such a long-wavelength instability qualitatively explains the two-step transition. The instability at a long wavelength naturally leads to a *long-range* velocity orientational correlation and, therefore, the fast increase of *M* at early times. However, since $E$ is determined by the integral of $E(k)$ over all $k$, the early-time increase of $E(k)$ at small $k$ does not substantially affect the total kinetic energy, which shows strong increase only when $E(k)$ at high $k$ starts to grow at later times. Hence, our measurements of transition kinetics provide experimental evidence confirming the key prediction of the kinetic theories on the existence of a long wavelength instability in suspension of pusher swimmers (*10-13*). The observation of the long incubation period near the phase boundary is, however, not captured by the kinetic theories. Although the growth rate of the long wavelength instability also approaches to zero near the phase boundary within the framework of the kinetic theories, one would expect to see the initial increase of $E(k)$ at small $k$ near the phase boundary. This prediction contradicts our *k*-space analysis (Fig. 5A), as well as the real-space observation shown in Fig. 3B, where the collective turbulent phase initiates locally at small scales and coalesces into large-scale structures only after the prolonged incubation period. Beyond the kinetic theories, the presence of two different transition kinetics is reminiscent of a theoretical phase diagram on the motility-induced phase separation of active fluids, where binodal and spinodal regions have been predicted based on the concept of swim pressures (*49*).

**Microscopic view.** Lastly, we examine the evolution of the long-wavelength instability from the perspective of the dynamics of single bacteria using high-resolution fast confocal microscopy (Figs. 6A, B) (Materials and methods). We find that the two-step transition deep inside the turbulent phase can be understood microscopically from the development of the orientational order of bacteria in emerging turbulent flows. Directly tracking the orientation of individual bacteria in dense 3D suspensions is difficult subject to large experimental errors. In order to quantify the local nematic order of bacteria without tracking individual bacteria, we perform Fast Fourier Transform (FFT) on local regions (20 × 20 μm$^2$) of confocal images. The degree of the local nematic order can be characterized by the anisotropy of the resulting FFT patterns, *F* (Fig. 6B inset) (Materials and methods). Our measurements show that the local bacterial nematic order is linearly proportional to the kinetic energy $E$, but decouples from the velocity orientational order $M$ in the two-step transition deep inside the turbulent phase (Figs. 6C, D).

The finding suggests the following physical picture. At early times, a slight deviation from the isotropic distribution of bacterial orientations induced by the long wavelength instability is sufficient to establish a strong velocity orientational correlation, where bacteria still appear to orientate randomly in the emergent turbulent flow as shown in Fig. 6A. The picture further corroborates the exceptional robustness of bacterial turbulence identified at large $f$ of the phase diagram (Fig. 2D). Over time, the kinetic energy gradually increases as the local bacterial alignment is enhanced by local coherent flows (Fig. 6C). This positive feedback loop between kinetic energy and bacterial alignment qualitatively agrees with the numerical solution of the kinetic equations in the nonlinear regime (*11*), where the alignment of pusher swimmers with the direction of local flows develops progressively through the transition and the increasing alignment in turn enhances the velocity of local flows. A strong steady-state turbulence is finally established when bacteria well align with their neighbors as shown in Fig. 6B, which gives rise to both a large kinetic energy and a high local bacterial nematic order. Thus, our confocal measurements provide a microscopic view of the kinetics of the long-wavelength-instability-driven two-step transition: the disordered phase (low $M$, low $E$, low $F$) $\rightarrow$ the disordered turbulent phase (high $M$, low $E$, low $F$) $\rightarrow$ the ordered turbulent phase (high $M$, high $E$, high $F$).

## Discussion and conclusion

We experimentally studied the emergence of the collective motion of bacterial suspensions, the so-called bacterial turbulence. By using three different *E. coli* strains and examining over a thousand of bacterial suspensions at different concentrations, swimming speeds and fractions of active swimmers, we mapped a phase diagram of 3D bulk bacterial flows over a large control parameter space in a single experimental system. Furthermore, taking the advantage of genetically engineered light-powered *E. coli* whose locomotion can be reversibly controlled by light, we imaged the onset of bacterial turbulence and conducted measurements on the kinetic pathways of bacterial turbulence transition.

The contribution of our study is two-fold. On the one hand, our experiments validated the assumption of the kinetic theories of collective bacterial motions and verified the main predictions of the theories on the transition point and the mode of instabilities. Specifically, the phase boundary of our experimental phase diagram showed a quantitative agreement with the prediction of the theories without fitting parameters. Our kinetic measurements further showed the existence of the long wavelength instability deep inside the turbulent phase and therefore provided an experimental support to this key theoretical prediction. In addition, our microscopic measurements on single bacterial dynamics through the long-wavelength instability revealed the rise of local bacterial nematic order and its coupling with local coherent flows, qualitatively testifying the numerical solution of the kinetic equations in the nonlinear regime.

On the other hand, our measurements also revealed new features of bacterial turbulent transition, unexpected from existing theories. Particularly, we explored the effect of doping immobile bacteria on the phase dynamics of bacterial flows, which allowed us to identify the dominant role of hydrodynamic interactions in the formation of bacterial collective swimming in 3D suspensions. The finding not only resolved the controversy over the relative importance of hydrodynamic versus steric interactions, but also illustrated the remarkable robustness of bacterial turbulent flows with high fractions of malfunctioning units. More importantly, our experiments uncovered two distinct kinetic pathways towards bacterial turbulence at different locales of the phase diagram, i.e., the one-step transition with long incubation periods near the phase boundary and the two-step transition driven by the long wavelength instability deep inside the turbulent phase. These two pathways exhibited drastically different transition rates and transient structures, suggesting two qualitatively different transition mechanisms.

Finally, our study also opens new questions for future experimental and theoretical development. First, it is unclear on the microscopic origin of the long incubation period in the one-step transition near the phase boundary. Our preliminary study indicates that the incubation process may be influenced by the formation of bacterial clusters in suspensions. Nevertheless, the incubation time can be significantly longer than the time scale associated with the dissolving of bacterial clusters close to the phase boundary, suggesting the generic nature of the phenomenon. To understand the nucleation and growth process with long incubation, prolonged experiments are required beyond the time scale of typical bacterial experiments, which pose an experimental challenge due to photobleaching and decreasing bacterial activity over time. How to incorporate the incubation process in the kinetic theories also presents a challenging theoretical question. Second, the long wavelength instability observed in our experiments indicates a system-size dependence of phase dynamics in confined systems. The change of the phase diagram and the transition kinetics with the system size needs to be further confirmed and measured in future experiments. Lastly, it is certainly worth of investigating the kinetic route to collective motions in other active fluids such as active cytoskeletons (*50*) and suspensions of colloidal swimmers (*51*) and checking to what extent the kinetic features observed in our experiments are generic for active fluids in general.

**Materials and methods**

**Bacteria.** Three strains of *E. coli* are used. (*i*) A wild-type *E. coli* K-12 strain (BW25113), which shows normal run-and-tumble motions with an average swimming speed $v$ = 28.7 μm/s (*43*). (*ii*) A "tumbler" strain, which is made by knocking out *cheZ* of a wild-type *E. coli* strain (RP1616). The Δ*cheZ* mutant shows constant tumbling motions. We use tumblers as passive particles. To deactivate the bacteria, we remove the flagella of tumblers via shear by pipetting tumbler suspensions through a narrow pipette tip a few tens of times. (*iii*) A light-powered *E. coli*, whose swimming speed can be reversibly controlled by light. We introduce a light-driven transmembrane proton pump, proteorhodopsin (PR), to wild-type *E. coli* (BW25113) by transforming the bacteria with plasmid pZE-PR encoding the SAR86 γ-proteobacterial PR-variant (*38*). The activity of PR is directly correlated with the light intensity. Thus, we can control the swimming speed of bacteria using light of different intensities. The quantitative relation between the light intensity (in terms of the power of the light source) and the average swimming speed of bacteria is measured by tracking the motion of bacteria in dilute suspensions (Fig. S1A). The response time of the bacteria is fast within a few seconds after the change of the light (Fig. S1B). Note that the highest speed of the light-powered bacteria at high light intensity is about 10 to 11 μm/s (Fig. 2C), which is smaller than that of the wild-type strain. Three strains share a similar body plan with the length of bacterial bodies $2a$ = 2.8 ± 0.4 μm and the radius of $b$ = 0.5 ± 0.2 μm based on direct imaging of bacteria, where the errors are the standard errors of the measurements.

All the three strains are cultured using a similar procedure. They are first cultured at 37.0 °C with a shaking speed at 250 rpm for 14-16 hours in terrific broth (TB) culture medium [trypotone 1.2% (w/v), yeast extract 2.4% (w/v), and glycerol 0.4% (v/v)]. The saturated culture is then diluted 1:100 in TB culture medium and grown at 30 °C for 6.5 hours. PR expression is triggered by 1 mM isopropyl *β*-D-thiogalactoside and 10 μM methanolic all-trans-retinal, which are supplemented in the mid-log phase for the synthesis of proteorhodopsin. The bacteria are harvested by gentle centrifugation (800*g* for 5 min) in the late log phase. After discarding the supernatant, we resuspend bacteria with motility buffer MB (0.01 M potassium phosphate, 0.067 M NaCl, 10 M EDTA, PH 7.0). The suspension is finally washed twice and adjusted to the target volume fraction. Note that the volume fraction of bacteria at the standard concentration of $n_0$ is $\phi$ = 0.0018, where $n_0 = 8 \times 10^8$ cells/ml is the *E. coli* concentration at OD$_{600}$ = 1.0. To create the mixture of tumblers and swimmers, we separately prepare suspensions of tumblers and swimmers both at the targeted volume fraction. We then mix the two suspensions at a predetermined ratio to control the fraction of active swimmers $f$.

**Sample preparation and video microscopy.** For the wild-type bacteria, oxygen is necessary to maintain the swimming of bacteria. We deposit a 2 μl wild-type bacterial suspension on a microscope coverslip, which forms a free suspension-air interface on the top. The droplet is millimetric in the lateral directions (*x-y*) normal to the imaging plane and about 150 μm in height (*z*). The coverslip and the suspension are further enclosed in a humid chamber of ~ 1000 mm$^3$ to reduce evaporation as well as perturbation due to ambient air flow. An inverted microscope is used to image bacterial flow 70 μm above the coverslip, where the effect of oxygen gradients is weak and bacterial activity is uniform. No system-wide persistent flow associated with aerotaxis is observed in our experiments. The microscopy videos show bacterial flow in the 2D *x-y* plane inside 3D suspensions (Movie S1).

For the light-powered bacteria, it is important to shut down the metabolic pathway of aerobic respiration, so that the locomotion of bacteria is solely controlled by the PR pump. We inject a suspension of light-powered bacteria into a sealed cell of 18 × 3 × 0.17 mm$^3$. For concentrated bacterial suspensions above $\phi$ = 0.062, bacteria stop swimming after a few minutes in the cell due to the depletion of oxygen. Microscope illumination is then used to switch on/off and control the activity of the PR pump for bacterial swimming. 2D bacterial flow is imaged 80 μm above the bottom wall.

For most of our experiments, 2D flow fields are imaged through an inverted bright-field microscope using a 10× (NA 0.3), 20× (NA 0.5) or 40× (NA 0.6) objectives. The field of view of images ranges from 640 × 640 μm$^2$ to 160 × 160 μm$^2$, depending on the lens used. To measure the velocity orientational order and the kinetic energy in steady states, one-minute videos are taken 15 minutes after wild-type *E. coli* samples are loaded or 2 minutes after a new light condition is imposed to light-powered *E. coli* samples. These times are sufficient for the sample to reach the steady state. Five-minute videos are taken for transition kinetics after light ramping. All the videos are recorded at 30 frames per second by a sCMOS camera. To control bacterial velocity, light intensity is tuned by the voltage of the light source. Three to twelve independent measurements are taken for each set of control parameters of our experiments.

To measure the dynamics of individual bacteria through the turbulent transition, we image fluorescent-labeled *E. coli* with an inverted fast confocal microscope using a 60× objective (NA 1.4). The green fluorescent protein expressed in bacteria is excited by a 488 nm laser. The field of view of the images is 180 × 120 μm$^2$. Bacterial flow is imaged 10 μm above the coverslip. Five-minute videos are taken at 10 frames per second.

**Image processing and data analysis.** The velocity field of bacterial suspensions is obtained from raw videos using standard Particle Imaging Velocimetry (PIV). For each pair of neighboring frames, the interrogation window size shrinks from 19.2 × 19.2 μm$^2$ to 4.8 × 4.8 μm$^2$ in three iterations (*52*). The final lattice spacing of the velocity field is 2.4 μm. The velocity field in Fig. 1A shows the velocity vectors of every other lattices. To verify the accuracy of PIV, we have also measured the dynamics of bacterial suspensions using Particle Tracking Velocimetry (PTV) on passive colloidal tracers embedded in bacterial suspensions. The PTV results qualitatively agree with the PIV findings on both the phase boundary and transition kinetics.

Energy spectra quantify the energy distribution over different length scales, $\lambda = 2\pi/k$, where $k$ is the wavenumber. To obtain energy spectra, we first calculate the Fourier transform of the 2D velocity field $v_x(x,y)$ and $v_y(x,y)$ to obtain $u_k(k_x,k_y)$ and $v_k(k_x,k_y)$. The point-wise kinetic energy density in the $k$-space is then computed $E(k_x, k_y) = \langle u_k(k_x,k_y)u_k^*(k_x,k_y) + v_k(k_x,k_y)v_k^*(k_x,k_y)\rangle/2$, where * represents the complex conjugate. Finally, the energy spectrum $E(k)$ is obtained by summing up $E(k_x,k_y)$ at a constant $k = (k_x^2 + k_y^2)^{1/2}$. An alternative way to calculate $E(k)$ through the Fourier transform of the two-point velocity correlation function $\langle \vec{v}(\vec{r}_0) \cdot \vec{v}(\vec{r}_0 + \vec{r})\rangle_{\vec{r}_0}$ yields quantitatively similar results.

To extract bacterial nematic order, we perform Fast Fourier transformation (FFT) on the local regions (20 × 20 μm$^2$) of confocal images. The FFT patterns are first smoothed with 3 neighboring pixels (0.15 μm$^{-1}$) and then fitted with ellipses. The aspect ratio of the ellipses, *a*/*b*, quantifies the local nematic order of bacteria (Fig. 6B inset). The average of the aspect ratio is finally taken over the whole field of view, $F = \langle a/b \rangle$. To calibrate our method, we show numerically that the anisotropy of the FFT pattern of local regions, *a*/*b*, is approximately linearly proportional to the local nematic order parameter of bacteria $S = \frac{1}{N}\sum_{j=1}^{N}(3\cos^2\alpha_j - 1)/2$ within the range of our experiments (Fig. S5), where $\alpha_j$ is the angle between the orientation of bacterium *j* with respect to the mean orientation of all the *N* bacteria in a local region in 3D.

**Acknowledgements**

**General:** We thanks Yi-Shu Tai, Kechun Zhang, Seunghwan Shin, Xinliang Xu, Kevin Dorfman and Stefano Martiniani for fruitful discussions and help with experiments and data analysis.

**Funding:** The research is supported by NSF CBET-1702352, CBET-2028652 and Packard Foundation.

**Author contributions:** Y.P. and X.C. designed the research and wrote the paper. Y.P. and Z.L. performed experiments. Y.P., Z.L. and X.C. analyzed data.

**Competing interests:** The authors declare no competing interests.

**Data and materials availability:** All data needed to evaluate the conclusions in the paper are present in the paper and/or the Supplementary Materials. Additional data related to this paper may be requested from the authors.

**Supplementary Materials**

Supplementary Material and Supplementary Videos for this article are available.

**Figures**

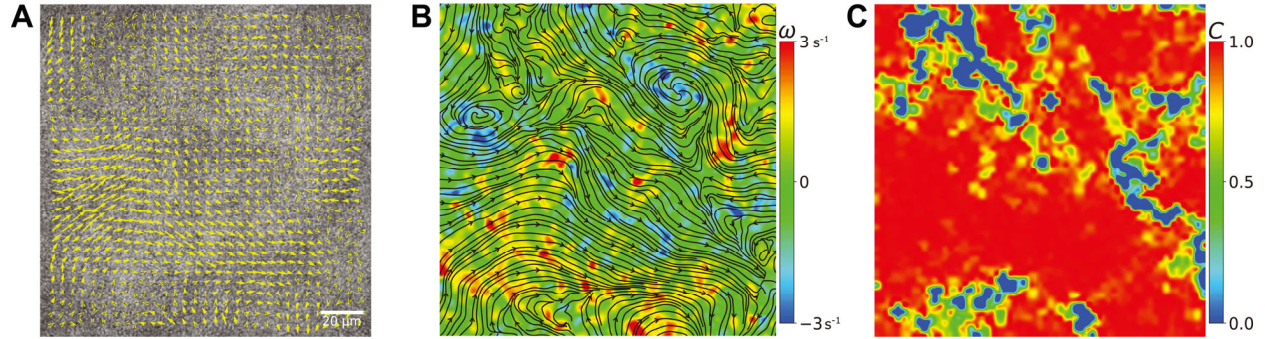

**Fig. 1 Bacterial turbulence.** (**A**) The flow field of a suspension of wild-type *E. coli* at volume fraction $\phi$ = 0.14. The arrows indicate local velocities from PIV. Scale bar: 20 μm. (**B**) The corresponding streamlines and the vorticity field ($\omega$), highlighting the characteristic features of bacterial turbulence including intermittent jets and vortices (see also Movie S1). (**C**) The corresponding velocity orientational correlation field *c*. The area fraction of red regions with $c \geq 0.9$, *M*, is used to quantify the transition to bacterial turbulence. $M < 1$ even for this highly turbulent state due to the presence of vortex cores and interstices.

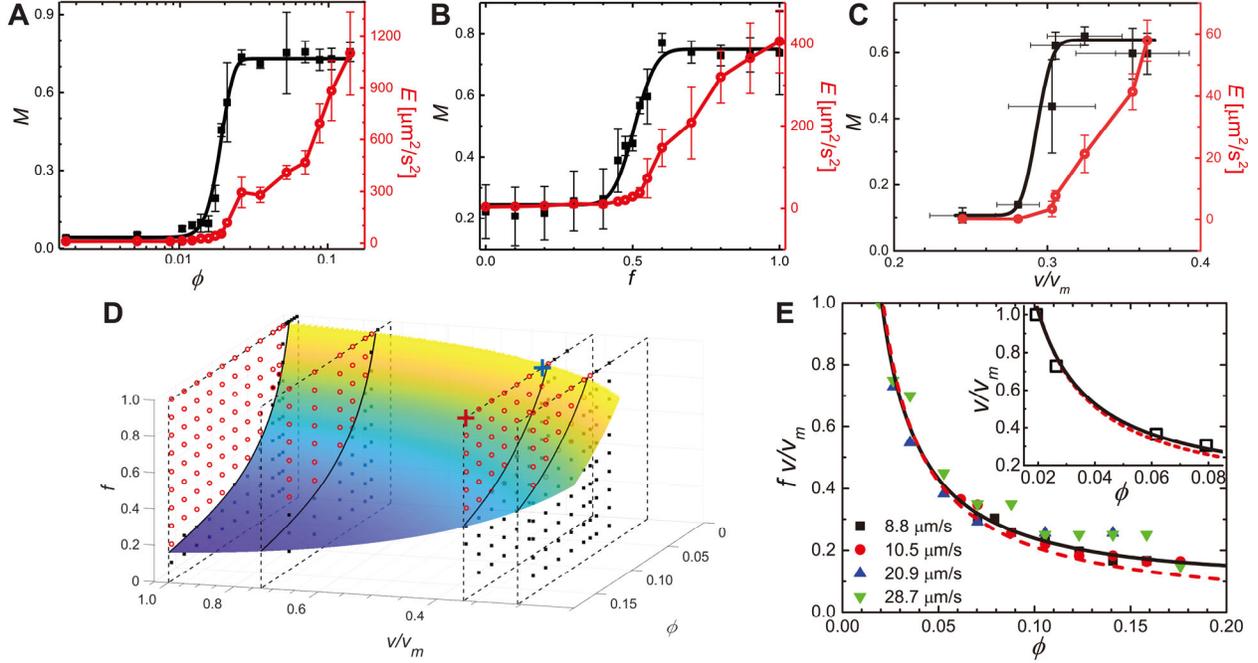

**Fig. 2 Transition to bacterial turbulence and the phase diagram of 3D bacterial flows.** (**A**) Velocity orientational order $M$ and kinetic energy $E$ as a function of bacterial volume fraction $\phi$ of wild-type bacteria with swimming speed $v_m$ = 28.7 μm/s and swimmer fraction $f$ = 1. Black squares are for $M$ and red circles are for $E$. (**B**) $M$ and $E$ versus $f$ at fixed $\phi$ = 0.053 and $v = v_m$. (**C**) $M$ and $E$ versus $v$ at $\phi$ = 0.11 and $f$ = 1. $v$ is normalized by the swimming speed of wild-type bacteria $v_m$. Black lines indicate error function fittings. Each data point in (A)-(C) represent one experiment averaged over 30 s in time. The error bars are the standard deviations of the temporal fluctuations of measured quantities. The error bars in $v$ are the standard deviations of the natural variation of bacterial swimming speeds in the dilute limit. Although for clarify we only show one set of experiments for each case, we have run more than 20 independent experiments near the phase boundary to locate the transition points. (**D**) A phase diagram of bacterial flows in the phase space spanned by $\phi$, $v$ and $f$. Red circles indicate the turbulent phase, whereas black squares indicate the disordered phase. The surface is the model prediction (Eq. (1)). The crosses indicate the locales of the kinetic experiments shown in Fig. 3 and 5. (**E**) The rescaled phase boundary between the disordered and the turbulent phases at different $v$ (indicated in the figure). The black line is the prediction of the simple hydrodynamic model (Eq. (1)), whereas the red dashed line is the prediction of the kinetic theory (Eq. (2)). Inset shows the experimental phase boundary and the theoretical predictions at $f$ = 1.

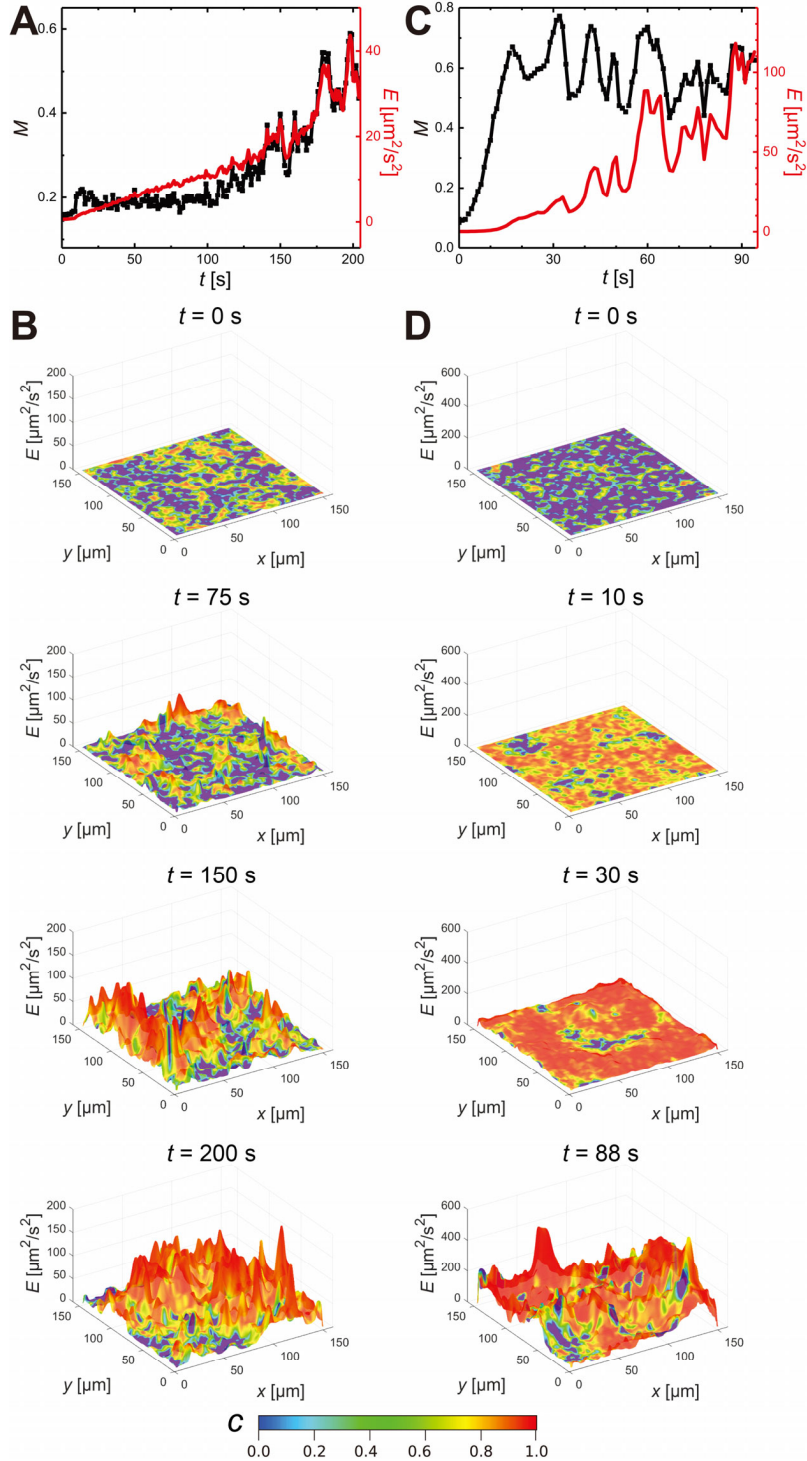

**Fig. 3 Transition kinetics.** (**A**) Velocity orientational order $M$ and kinetic energy $E$ as a function of time through the turbulent transition. $t = 0$ corresponds to the time when light intensity is ramped up. Bacterial volume fraction $\phi = 0.071$, speed $v = 10.5$ μm/s and swimmer fraction $f = 1$, which is near the phase boundary shown by the blue cross in Fig. 2D (see also Movie S2). (**B**) Spatiotemporal evolution of the local

velocity orientational correlation $c$ and the local kinetic energy $E$ over the transition shown in (**A**). The spatial coordinates are shown in the *x-y* plane. The height shows the value of $E$, whereas the color indicates the value of $c$. (**C**) $M$ and $E$ as a function of $t$ for $\phi = 0.18$, $v = 10.5$ μm/s and $f = 1$ deep inside the turbulent phase shown by the red cross in Fig. 2D (see also Movie S3). (**D**) Spatiotemporal evolution of $c$ and $E$ over the transition shown in (**C**).

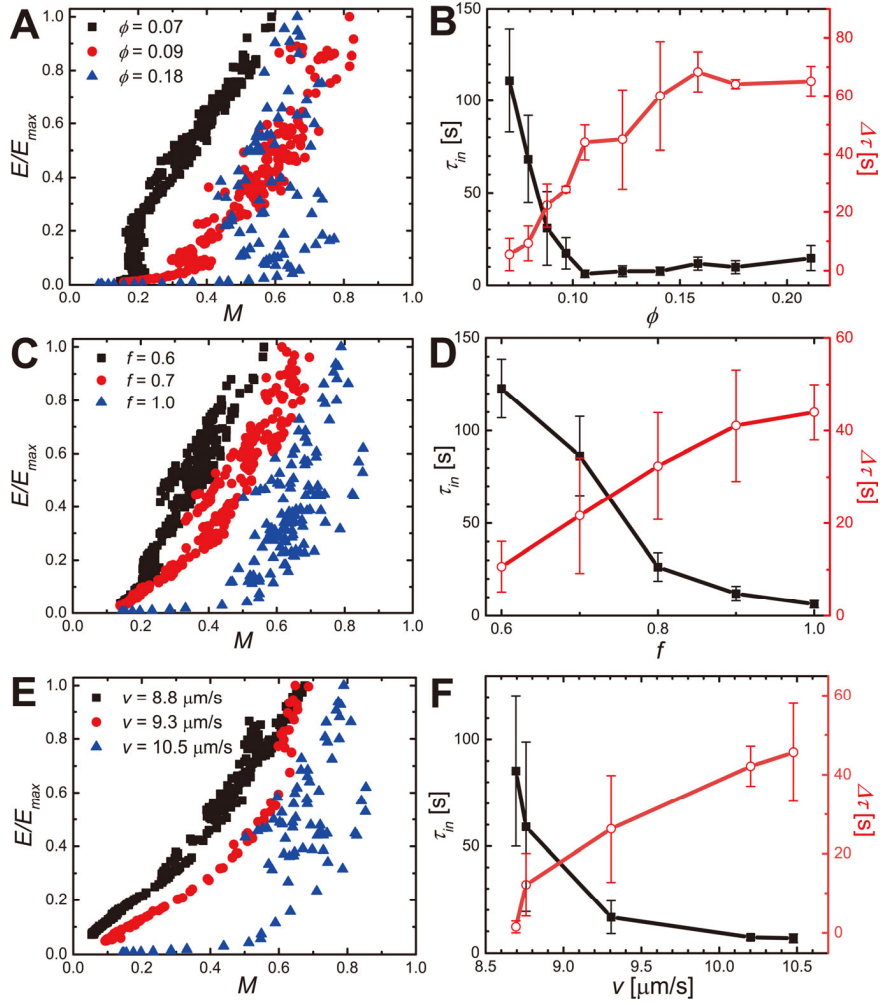

**Fig. 4 One-step and two-step transitions.** (**A**) Kinetic energy $E$ versus velocity orientational order $M$ over transitions at different bacterial volume fractions $\phi$. Bacterial speed and swimmer fraction are fixed at $v =$ 10.5 μm/s and $f = 1$, respectively. While the one-step transition shows an approximately linear concurrent increase of $E$ and $M$ over time (black squares), the two-step transition shows an increase $M$ at low $E$ in the first step at early times and then an increase of $E$ at an approximately constant $M$ in the second step at later times (blue triangles). $E$ is normalized by the maximal energy density at the steady state, $E_{max}$. (**B**) Incubation time, $\tau_{in}$, and the time difference, $\Delta\tau$, as a function of $\phi$ at the same $v$ and $f$ as in (**A**). The error bars reflect the standard error of three to twelve experimental runs for each data point. (**C**) $E/E_{max}$ versus $M$ at different $f$ and fixed $\phi = 0.11$ and $v = 10.5$ μm/s. (**D**) $\tau_{in}$ and $\Delta\tau$ at different $f$ at the same $\phi$ and $v$ as in (**C**). (**E**) $E/E_{max}$ versus $M$ at different $v$ and fixed $\phi = 0.11$ and $f = 1$. (**F**) $\tau_{in}$ and $\Delta\tau$ at different $v$ at the same $\phi$ and $f$ as in (**E**).

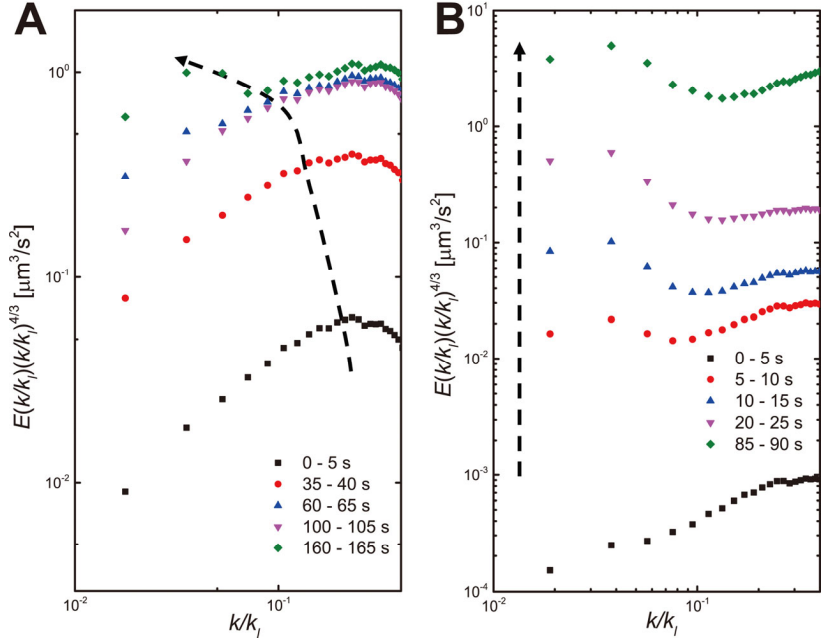

**Fig. 5 Energy spectra through the bacterial turbulent transition.** (**A**) Temporal evolution of energy spectrum $E(k)$ through the turbulent transition for a system near the phase boundary (blue cross in Fig. 2D). Bacterial volume fraction $\phi = 0.071$, speed $v = 10.5$ μm/s and swimmer fraction $f = 1$. (**B**) Temporal evolution of $E(k)$ through the turbulent transition for a system deep inside the turbulent phase (red cross in Fig. 2D). $\phi = 0.18$, $v = 10.5$ μm/s and $f = 1$. Since $E(k)$ shows a $k^{-4/3}$ scaling at high $k$ from direct fitting, we plot $E(k)(k/k_l)^{4/3}$ to make the curves flat at high $k$, where $k_l = 2\pi/3$ μm$^{-1}$ is the wavenumber for the size of single bacteria. The dashed arrows indicate different increasing trends of $E(k)$.

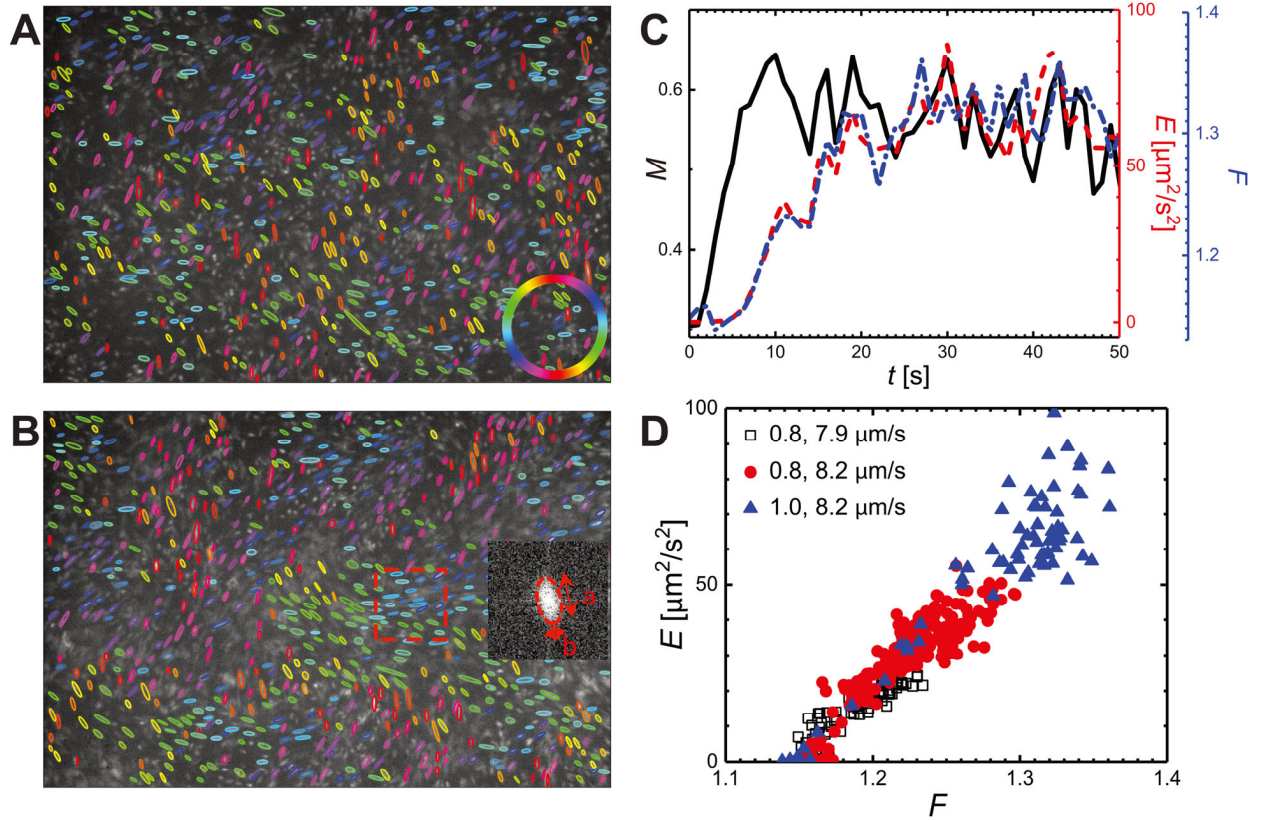

**Fig. 6 Microscopic view of the turbulent transition deep inside the turbulent phase.** Confocal microscopy of a bacterial suspension at $t = 10$ s (**A**) and 30 s (**B**) after the light ramp. The incubation time of the transition $\tau_{in} \approx 6$ s. Bacterial volume fraction $\phi = 0.14$, speed $v = 8.2$ μm/s and swimmer fraction $f = 1$. Individual bacteria identified manually are marked with ellipsoids, whose orientation is indicated by the color bar in the inset of (**A**). The inset of (**B**) shows the FFT of a local region of $20 \times 20$ μm$^2$ indicated by the dashed box. The aspect ratio of FFT patterns averaged over all the local regions, $F \equiv \langle a/b \rangle$, is used to quantify the local nematic order of bacteria (Materials and methods). (**C**) Velocity orientational order, $M$, kinetic energy, $E$, and $F$ as a function of time $t$ through the transition shown in (**A**) and (**B**). The black solid line is for $M$ shown to the left axis, whereas the red dashed line and the blue dash-dotted line are for $E$ and $F$, respectively, shown to the right axis. $E$ and $F$ show quantitatively similar trends, different from that of $M$. (**D**) $E$ versus $F$ at $\phi = 0.14$ and different $f$ and $v$ (indicated in the figure). The open squares are for a system close to the phase boundary with the one-step transition, whereas the blue triangles are for a system deep inside the turbulent phase with the two-step transition. The red disks are for a crossover system between the two limits. Different from the $E$-$M$ plots shown in Fig. 4A, C and E, all the data collapse into a master curve showing the linear relation between $E$ and $F$.